\documentclass[doc, floatsintext]{apa6}
\usepackage{apacite}

\usepackage{amsmath,amssymb,amsthm,enumitem}
\usepackage{setspace}
\usepackage{amsmath}
\usepackage{graphicx}
\usepackage[]{color}
\usepackage{epsfig}
\usepackage{latexsym}
\usepackage{amsmath}
\usepackage{amsthm}
\usepackage{graphicx}
\usepackage{amssymb}

\newtheorem{proposition}{Proposition}

\theoremstyle{definition}

\newcommand{\G}{\mathcal{G}}

\newcommand{\D}{\mathcal{D}}

\newcommand{\cR}{\mathcal{R}}
\newcommand{\cQ}{\mathcal{Q}}

\newcommand{\cC}{{\mathcal C}}
\newcommand\ring[1]{\mathaccent23{#1}}

\newcommand{\citep}{\cite}

\newlength{\normalparindent}
\raggedright
\doublespace

\doublespace

\AtBeginEnvironment{tabular}{\doublespacing}

\title{Modeling a Cognitive Transition at the Origin of Cultural Evolution using Autocatalytic Networks}
\twoauthors{Liane Gabora}{Mike Steel}
\twoaffiliations{Department of Psychology, University of British Columbia}{Biomathematics Research Centre, University of Canterbury}  

\shorttitle{COGNITIVE TRANSITION CULTURE AUTOCATALYTIC NETWORKS}

\authornote{LG acknowledges funding from Grant 62R06523 from the Natural Sciences and Engineering Research Council of Canada. We thank Russell Gray for drawing our attention to the paper by Stout et al. that inspired this one. We thank Claes Andersson and anonymous reviewers for comments, and thank Conner Gibbs for assistance with the manuscript.}

\begin{document}
\doublespace 

\noindent Running Head: \\ \noindent COGNITIVE TRANSITION CULTURE AUTOCATALYTIC NETWORKS
\maketitle

\bigskip
\noindent This is a pre-publication draft. There may be minor differences from the version accepted
for publication in Cognitive Science.\\
\bigskip

\bigskip

\noindent Corresponding Author: \\
\noindent Liane Gabora \\
\noindent Department of Psychology, University of British Columbia \\
\noindent Okanagan Campus, Kelowna BC, Canada
\\ 
\noindent Email: liane.gabora@ubc.ca

\newpage

\begin{center}
{\bf Abstract}
\end{center}

Autocatalytic networks have been used to model the emergence of self-organizing structure capable of sustaining life and undergoing biological evolution. Here, we model the emergence of {\it cognitive} structure capable of undergoing {\it cultural} evolution. Mental representations of knowledge and experiences play the role of catalytic molecules, and interactions amongst them (e.g., the forging of new associations) play the role of reactions, and result in representational redescription.
The approach tags mental representations with their source, i.e., whether they were acquired through social learning, individual learning (of {\it pre-existing} information), or creative thought (resulting in the generation of {\it new} information). This makes it possible to model how cognitive structure emerges, and to trace lineages of cumulative culture step by step. We develop a formal representation of the cultural transition from Oldowan to Acheulean tool technology using Reflexively Autocatalytifc and Food set generated (RAF) networks. Unlike more primitive Oldowan stone tools, the Acheulean hand axe required not only the capacity to envision and bring into being something that did not yet exist, but hierarchically structured thought and action, and the generation of new mental representations: the concepts EDGING, THINNING, SHAPING, and a meta-concept, HAND AXE. We show how this constituted a key transition towards the emergence of semantic networks that were self-organizing, self-sustaining, and autocatalytic, and discuss how such networks replicated through social interaction. The model provides a promising approach to unraveling one of the greatest anthropological mysteries: that of why development of the Acheulean hand axe was followed by over a million years of cultural stasis.\\

\bigskip


\newpage

\section{Introduction}
The question of how biological evolution arose---i.e., the origin of life (OOL) problem---is one of the biggest unsolved questions of science. Since cultural change is widely viewed as a second evolutionary process, the question of how {\it cultural} evolution arose---i.e., the origin of culture (OOC) problem---presents another unsolved problem. By {\it culture}, we mean extrasomatic adaptations, including behavior and artifacts, that are socially rather than genetically transmitted.
Although cultural {\it transmission}---in which one individual acquires elements of culture from another---is observed in many species, cultural {\it evolution} is much rarer (and perhaps, unique to our species).\footnote{The term `cultural evolution' is occasionally used in a less restricted sense to refer to novelty generation and transmission without the requirement of cumulative, adaptive, open-ended change, e.g., \cite{whi19}.} 
By {\it evolution}, we mean change that is cumulative (later innovations build on earlier ones), adaptive new innovations that yield some benefit for their bearers tend to predominate), and open-ended (the space of possible innovations is not finite, since each innovation can give rise to spin-offs).
The literature on cultural evolution, including mathematical and computational models, is vast, flourishing, and interdisciplinary \cite{benetal04,boyric88,cavfel81,endetal11,gab13,holmac03,meswhilal06,powetal09} with increasing recognition paid to the cognitive processes and abilities (e.g., problem solving, analogy, and so forth) underlying the generation of cultural novelty \cite{fog15,gab19,hen20,hey18,ove19}. 
However, although cognitive science has made considerable progress in understanding how such processes are carried out, little effort has been devoted to the question of how hominids acquired the capacity for an integrated conceptual framework that guides how and when these processes are applied.

This paper addresses what kind of structure minds must possess to be capable of cumulative, open-ended cultural evolution, and how hominid minds acquired this kind of structure. We propose a network-based model that tags mental representations with their source, i.e., whether they were acquired through individual learning, social learning, or creative reflection. This makes it possible to model how a semantic network emerges, and to trace cumulative change in cultural lineages step by step. The approach is demonstrated using a formal representation of one of the earliest and most well-studied transitions in human cultural history: the transition from Oldowan to Acheulean tool technology approximately 1.76 million years ago (mya).

Although evolutionary theory is widely applied to culture, natural selection cannot shed light on the {\it origins} of an evolutionary process (as Darwin himself noted); it can only explain how, once self-sustaining, self-reproducing entities have come into existence, they evolve.\footnote{Research on epigenetics and the origin of life has shown that natural selection is but one (albeit important) component of evolution \cite{gab06,kau1,koo09,seg00,vet06,woe02}.}  However, although natural selection does not address the `origins' question, another theory, the theory of autocatalytic networks, does. 
This theory grew out of studies of the statistical properties of {\it random graphs} consisting of nodes randomly connected by edges \cite{erdosrenyi1960}. As the ratio of edges to nodes increases, the size of the largest cluster increases, and the probability of a phase transition resulting in a single giant connected cluster also increases.  
The recognition that connected graphs exhibit phase transitions led to their application to efforts to develop a formal model of the OOL, namely, of how abiogenic catalytic molecules crossed the threshold to the kind of collectively self-sustaining, self-replicating, evolving structure we call `alive' \cite{kau1, kau2}.
In the application of graph theory to the OOL, nodes represent catalytic molecules, and edges represent reactions. 
It is exceedingly improbable that any catalytic molecule present in the primordial soup of Earth's early atmosphere catalyzed its own formation. However, reactions generate new molecules that catalyze new reactions, and as the variety of molecules increases, the  variety of reactions increases faster. As the ratio of reactions to molecules increases, the probability increases that they will undergo a phase transition. When, for each molecule, there is a catalytic pathway to its formation, they are said to be collectively {\it autocatalytic}, and the process by which this state is achieved has been referred to as {\it autocatalytic closure} \cite{kau1}.
The molecules thereby become a self-sustaining, self-replicating  structure (i.e., a living protocell \citep{hord15}). Thus, the theory of autocatalytic networks has provided a promising avenue for modeling the OOL and thereby understanding how biological evolution began \cite{xav20}. 

Autocatalytic networks have been developed mathematically in the theory of Reflexively Autocatalytic and Food set generated (RAF) networks \cite{hor16,ste19}. 
The term {\it reflexively} is used here in its mathematical sense, meaning that every element is related to the whole. The term {\it food set} refers to the reactants that are initially present, as opposed to those that are the products of catalytic reactions. It has been demonstrated (both in theory and in simulation studies) that RAFs can evolve (through selection and drift acting on possible subRAFs of the maxRAF) \cite{hor16,vas12}.

It has been proposed that autocatalytic networks hold the key to understanding the origins of {\it any} evolutionary process, including the OOC \cite{gab98,gab00,gab13,gabaer09,gabste}.\footnote{For related approaches, see \cite{andtor19,cab13,mut18}.} This kind of evolution has been referred to as Self-Other Reorganization (SOR) because it interleaves internal self-organization with external interactions with other self-organizing networks \cite{gab19}.
In application to the OOC, the products and reactants are not catalytic molecules but {\it mental representations}
{\footnote{Although we use the term `mental representation,' our model is consistent with the view (common amongst ecological psychologists and in the situated cognition and quantum cognition communities) that what we call mental representations do not `represent,' but rather, act as  contextually elicited bridges between mind and world.} 
(MRs) of experiences, ideas, and chunks of knowledge, as well as more complex mental structures such as schemas and scripts (Tables~\ref{gt} and~\ref{abbrev}). 

MRs are composed of one or more {\em concepts:} mental constructs such as ISLAND or BEAUTY that enable us to interpret new situations in terms of similar previous ones.
The rationale for treating MRs as catalysts comes from the literature on concept combination, which provides extensive evidence that when concepts act as contexts for other concepts, their meanings change in ways that are often nontrivial and that defy classical logic \cite{aeretal16,aeretal13,ham88,oshsmi81}. The extent to which one MR modifies the meaning of another is referred to here as its {\em reactivity}. A MR's reactivity varies in a context-sensitive manner. For example, in a study of the influence of context and mode of thought on the perceived meanings of concepts (as measured by property applicabilities and exemplar typicalities), the concept PYLON was rated low as an exemplar of HAT, but in the context FUNNY (as in `worn to be funny') it was rated high as an exemplar of HAT \cite{vel11}. Thus, the degree to which PYLON qualified as an instance of a HAT changed dramatically depending on the context. The context FUNNY had an even greater effect on the rating of MEDICINE HAT (as in the name of the Canadian town) as an instance of HAT. We say that the {\it reactivity} was high here because the context exerted a dramatic influence on the perceived meaning. Each interaction between two or more MRs alters (however slightly) the network of association strengths in memory \cite{bro10,mcc11}.
Eventually, for each MR, there is an associative pathway to its  formation, i.e., any given concept can be explained using other concepts, and new ideas can be reframed in terms of existing ones.
 
 \bigskip

\begin{table}[ht]
\caption{Application of graph theoretic concepts to the origin of life (OOL) and origin of culture (OOC).}
\begin{center}
\begin{tabular}{@{} lll @{}}
\hline \hline 
\textbf{Graph Theory} & \textbf{Origin of Life (OOL)} & \textbf{Origin of Culture (OOC)}\\ 
\hline
node & catalytic molecule & mental representation (MR)\\
edge & reaction pathway & association \\
cluster & molecules connected via reactions & MRs connected via associations \\
connected graph & autocatalytic closure \cite{kau2,kau1} & conceptual closure\footnote{Conceptual closure is the focus of another paper (in progress).} \cite{gab98}\\
\hline
\hline
\end{tabular}
\end{center}
\label{gt}
\end{table}

\begin{table}[ht]
\caption{Abbreviated terms used throughout this paper.}
\begin{center}
\begin{tabular}{@{} lll @{}}
\hline \hline 
\textbf{Abbreviation} & \textbf{Meaning}\\ 
\hline
OOL   & Origin of Life \\
OOC  & Origin of Culture \\
MR   & Mental Representation \\
sMR & simple Mental Representation \\
cMR & complex Mental Representation \\
RR  & Representational Redescription \\
RAF & Reflexively Autocatalytic and Food set generated (F-generated)\\
CCP &  Cognitive Catalytic Process \\
\hline
\hline
\end{tabular}
\end{center}
\label{abbrev}
\end{table}

In previous work, we used the RAF framework to model an initial transition toward the kind of cognitive organization capable of evolving culture \cite{gabste}. Our model followed up on the proposal that the increased complexity of {\em Homo erectus} culture compared with other species such as {\em Homo habilis} reflected the onset of {\it representational redescription} (RR), in which the contents of working memory were recursively restructured by drawing upon similar or related ideas \cite{cor11,don91,gabsmi18,hau02,pen08}. We showed how the capacity for RR would have enabled the forging of associations between MRs, thereby constituting a key step toward an essentially `autocatalytic' structure \cite{gabste}. The present paper elaborates on the approach, showing how RR enabled the emergence of hierarchically structured concepts, making it possible to shift between levels of abstraction as needed to carry out tasks composed of multiple subgoals.

To address how the mind as a whole acquired autocatalytic structure, the model presented here is, by necessity, abstract. This paper does not distinguish between semantic memory (memory of words, concepts, propositions, and world knowledge) and episodic memory (personal experiences); indeed, we are sympathetic to the view that these are not as distinct as once thought \cite{kwa05}. Nor does it address how MRs are obtained (i.e., whether through Hebbian learning versus probabilistic inference). Although MRs are represented simply as points in an $N$--dimensional space (where $N$ is the number of distinguishable differences, i.e., ways in which MRs could differ), our model is consistent with models that use convolution \cite{JonMew07}, random indexing \cite{kan09}, or other methods of representing MRs. 
We assume that associations form between MRs but we do not address whether these associations are due to similarity or co-occurrence, and whether they are learned through Bayesian inference  \cite{grifetal07} or other means. 
We view associations as probabilistic; thus when we say that a new association has been forged between two concepts we mean a spike in the probability of one MR evoking another, which we refer to here as the `catalysis' of one MR by the other.
We view context as anything that influences the instantiation of a MR in working memory (for example, the properties it possesses, or the exemplars it instantiates). Context in our model can be either external (e.g., an object or person) or internal (e.g., other MRs). Although our approach is influenced by how context is modeled in quantum approaches to concepts \cite{aeretal13,aeretal16}, it is not committed to any formal approach to modeling context.

\section{Autocatalytic networks}


 

We now summarize the key concepts of RAF theory. 
A {\it catalytic reaction system} (CRS) is a tuple $\cQ=(X,\cR, C, F)$ consisting of a set $X$ of molecule types, a set $\cR$ of reactions, a catalysis set $C$ indicating which molecule types catalyze which reactions, and a subset $F$ of $X$ called the food set. 
A {\it Reflexively Autocatalytic and F-generated}  set---i.e., a RAF---is a non-empty subset $\cR' \subseteq \cR$ of reactions that satisfies the following two properties:
\begin{enumerate}
  \item {\it Reflexively autocatalytic}: each reaction $r \in \cR'$ is catalyzed by at least one molecule type that is either produced by $\cR'$ or is present in the food set $F$; and
  \item {\it F-generated}: all reactants in $\cR'$ can be generated from the food set $F$ by using a series of reactions only from $\cR'$ itself.
\end{enumerate}

A set of reactions that forms a RAF is simultaneously  self-sustaining (by the $F$-generated condition) and  (collectively) autocatalytic (by the RA condition) because each of its reactions is catalyzed by a molecule associated with the RAF. A CRS need not have a RAF, but when it does there is a unique maximal one. Moreover, a CRS, may contain many possible RAFs, and it is this feature that allows RAFs to evolve (as described in Section 6 of (Hordijk \& Steel, 2016); see also (Vasas et al., 2012)).\\
In the OOL context, a RAF emerges in systems of polymers (molecules consisting of repeated units called monomers) when the complexity of these polymers (as measured by maximum length) reaches a certain threshold \cite{kau1,mos}. The phase transition from no RAF to a RAF incorporating most or all of the molecules depends on (1) the probability of any one polymer catalyzing the reaction by which a given other polymer was formed, and (2) the maximum length (number of monomers) of polymers in the system. 
This transition has been formalized and analyzed (mathematically, and using simulations), and applied to real biochemical systems \cite{horetal10, horetal11, horsteel04, hor16, mos} and ecologies \cite{gat2}. RAF theory has proven useful for identifying how phase transitions might occur, and at what parameter values.\\
In this application of RAFs to the OOC, we first summarize the archaeological evidence for a cognitive transition approximately 1.7 mya, limiting the discussion to aspects that either were not addressed elsewhere \cite{gabste} or that are essential in order to follow the arguments presented here. We then discuss cognitive mechanisms underlying the invention of the Acheulean hand axe, and present the RAF model of this. We conclude with a discussion of the implications of this new approach and ideas for further developments.

\section{Evidence for a cognitive transition}

The large cranial capacity of {\em Homo erectus} (approximately 1000 cc, 25\% larger than that of {\em Homo habilis} \cite{aie96}) is believed to have played a role in a cultural transition to significantly more complex tools, as epitomized by the Acheulean hand axe 1.76 mya, shown in Fig.~\ref{oldowan-vs-acheulean} \cite{edw01}.
Like its predecessor the Oldowan stone flake, the Acheulean hand axe was a multi-use implement, but whereas the former simply requires repeated percussive action, the latter is notoriously difficult to make \cite{par19,sto08}. It requires a skilled, multi-step process involving multiple different hierarchically organized actions, as shown in Fig.~\ref{making-acheulean}.\footnote{Detailed comparison of the making of Oldowan ({\em Homo habilis}) versus Acheulean ({\em Homo erectus}) tools, including the brain regions involved, can be found in \cite{sto08}.} 

\begin{figure}
\centering
\fitfigure[scale=0.4]{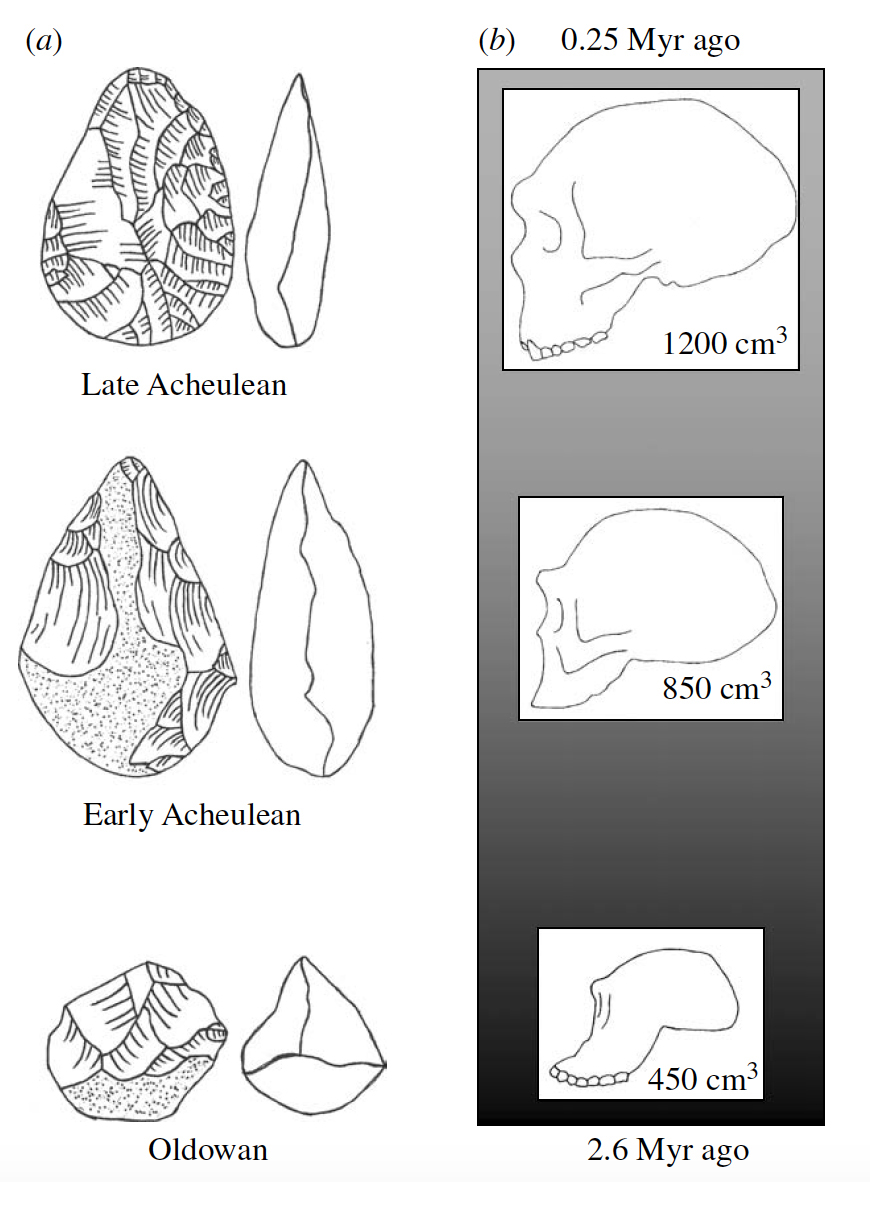} 
\caption{ Change in Early Stone Age technology and cranial capacity. From (Stout, 2008).}
\label{oldowan-vs-acheulean}
\end{figure}

The Acheulean hand axe is the most tangible evidence of a cognition transition characterized by a suite of related abilities. This transition is thought to have involved the onset of {\it autocuing}: the capacity to voluntarily retrieve a specific memory item in the absence of environmental cues \cite{don91}. Encephalization likely played a role in the onset of autocuing by enabling memories to be encoded in sufficient detail to evoke one another based on relevant (i.e., situation-specific) similarities \cite{gabsmi18}. Autocuing paved the way for {\it mental time travel}: the capacity to escape the immediate present by remembering past episodes or imagining events taking place at other locations, or in the future \cite{cor11}. Autocuing and mental time travel enabled individuals to engage in imaginative thought, and to reflect upon and update (i.e., elaborate, modify, restructure, and/or perform mental operations upon) the contents of working memory, drawing upon relevant knowledge or experience, as needed. This kind of recoding of information has been referred to as {\it representational redescription} (RR) \cite{kar92}. In this paper, the term RR refers to any internally-generated modification to a MR's network of associations, whether it be the result of a sudden flash of insight, or a new perspective on, or application for, an old idea. The capacity for autocuing, mental time travel, and RR culminated in a suite of related abilities that include the rehearsal and refinement of skills and the miming of past or possible future events \cite{cor11,don91}.

The recursive application of RR, such that the output of one redescription serves as the input to the next, we refer to as {\it abstract thought}. It may be accompanied by behavioral action that modifies the environment, and thereby tracks or externally manifests this internal process.
It is hypothesized that such MRs consisted of not just sensory representations of raw materials and the tools of which they are made, but also of their affordances with respect to the body, i.e., their capacity to be altered, used, or manipulated. (See \cite{ove19} for further discussion of the content of primitive toolmakers' MRs).
 
\begin{figure}
\centering
\fitfigure[scale=0.2]{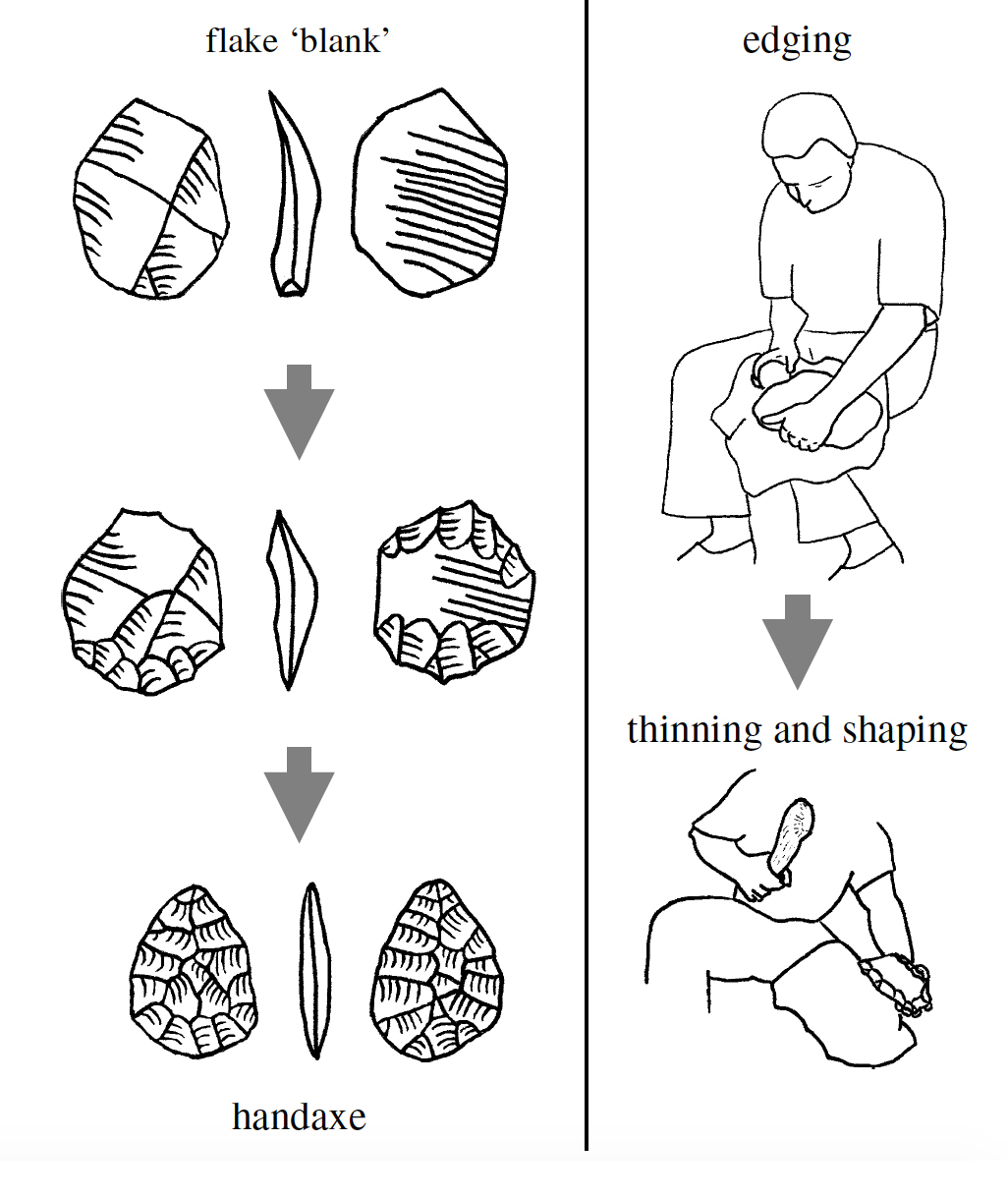}
\caption{Left: initial (top), intermediate (middle), and final (bottom) stages in the making of an Acheulean hand axe. Right: Physical processes required to bring the tool from initial to intermediate stage (top), and from the intermediate to the final stage (bottom). From (Stout, 2008).}
\label{making-acheulean}
\end{figure}

Summarizing an argument developed in detail elsewhere  (including discussions of the relationship of RR to the concept of `merge' \cite{gabsmi18}, and to the psychological literature on concept combination \cite{gabaer09,gabkit13,gabste}, we propose that RR was made possible because the larger {\em Homo erectus} brain enabled a finer-grained associative memory such that episodic and semantic knowledge could be encoded in greater detail, as illustrated in Fig.~\ref{nature-nurture}. 
This meant that, once an individual had lived lived long enough to accumulate a sufficiently rich repository of experience, there was more overlap in his or her distributed representations. This in turn meant that there were more routes by which these MRs could evoke one another, and more ways for the individual to generate novel cultural contributions. 

\begin{figure}
\centering
\includegraphics[scale=0.72]{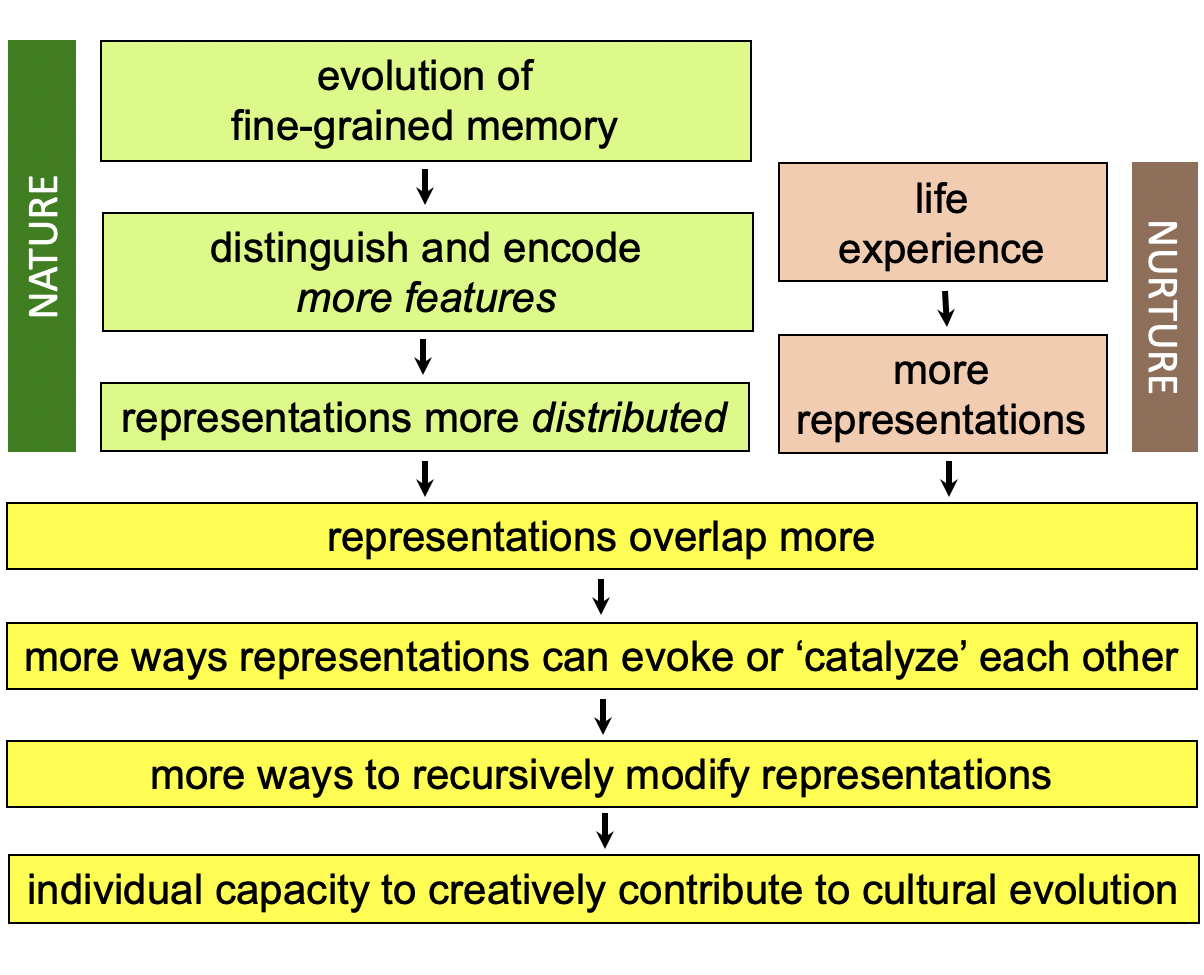}
\caption{Schematic illustration of the proposed changes set into motion through the evolution of finer grained memory, culminating in an individual's capacity to become a creative contributor to cultural evolution. Note that, for this sequence of events to unfold in a given individual, not only must that individual's representations be sufficiently distributed (due to the biologically evolved changes depicted on the upper left) but they must also be sufficiently plentiful (due to the developmental changes depicted on the upper right).}
\label{nature-nurture}
\end{figure}

RR enabled {\em Homo erectus} to combine MRs and chain them into streams of thought or sequences of action, and mime past or possible future event sequences to others (since it is generally believed that complex language was not yet established).\footnote{Although the evolution of language is stubbornly resistant to empirical inquiry \cite{hau14,per12}, it is thought that language as we know it, with syntax and grammar, came substantially after the Oldowan to Achuelian transition modeled here \cite{kle09,lie07,mit06}, most likely preceded by `protolanguage' \cite{bic07}, and the use of gesture and mime \cite{cor11,cor13,don91}. 
In this way, the content of experience could become detached from the immediate sensory perceptions of the `here and now' (in what has been called `mental time travel' \cite{cor11}).
Since RR forges new associations between MRs, the onset of the capacity for RR could, through spreading activation, affect other MRs, thereby facilitating categorization, generalization, and property induction. 
RR facilitated the formation of abstract concepts, including complex and sometimes spontaneously-generated {\em ad hoc} concepts 
\cite{bar83}. Abstraction may be facilitated by dimensional reduction, ensuring that concepts contain no more detail than necessary. The MR of an object could be redescribed as affording different actions and uses, so as to alternate between different subgoals, as needed, to reach an ultimate goal. Note that unlike purely social hypotheses regarding this transition, RR would facilitate not just the ability to {\it make} tools but the ability to demonstrate the process step-by-step to others; thus it could be put to use in social settings as well as technical tasks. The fact that PET imaging studies indicate that toolmaking and language share overlapping neural circuity \cite{sto08} suggests that RR may have contributed to the evolution of language.
We note that the model proposed in this paper does not rely on a precise timeline for language origins.}}

\subsection{Cognitive processes in Oldowan versus Acheulean toolmaking}

We now delve more deeply into the cognitive mechanisms underlying onset of the capacity for Acheulean toolmaking, arguably the earliest significant transition in the archaeological record.
We cannot know exactly how the first Oldowan tool was invented, and there are different scenarios by which it could have come about. 
One scenario is that the inventor of this tool imagined the impact of percussive action on a stone, and creatively redescribed the transformation of a stone into a tool. In this scenario, the new concept TOOL was not imported from the external world; rather, using RR, the toolmaker generated a mental image of something that did not currently exist. The hominin developed a {\it complex MR} (cMR) composed of the {\it simple MRs} (sMRs) STONE, TOOL, and PERCUSSIVE ACTION. This cMR can be understood as: in the context of the catalyst {\it percussive action}, STONE may become a TOOL. 

However, there are other scenarios by which this cMR could have come about, through individual learning. {\it Individual learning} refers to learning from the environment by nonsocial means through direct perception.
Note that in much of the cultural evolution literature, abstract thought and creativity, if mentioned at all, are equated with individual learning, which is thought to mean `learning for oneself' (e.g., \cite{henboy02,meswhilal06,rog88}. However, individual learning is distinctly different from abstract thought. In individual learning (obtaining pre-existing information from the environment through non-social means), the {\it information does not change} form just because the individual now knows it. Going into a forest and learning for oneself to distinguish different kinds of insects is an example of individual learning. In contrast, in abstract thought (reiterative processing of internally sourced mental contents), the {\it information is in flux} \cite{bar05}, and when this incremental honing process results in the generation of new and useful or pleasing ideas, behavior, or artifacts, it is said to be creative \cite{bas95,cha15,fei06,gab17}.

Since in individual learning, the information retains the form in which it was originally perceived, it does not involve RR.
For example, upon seeing boulders fall from a cliff and splinter a stone flake below into something that could be used as a tool, a hominid could have then imitated the percussive action of the falling boulders to create the first intentionally manufactured tool. The distinction between individual learning and RR is not black and white; it is possible that a certain amount of redescription was required to realize that one could intentionally mimic the action of the boulder on the rock.
However, applying Occam's razor, we will model the simple possibility that the initial idea for the cMR of using percussion to make a stone tool emerged through imitation of some kind of accidental breakage of a stone flake, and RR was not required. Note that even in this simple scenario, the toolmaking process was far from trivial; it required careful deliberate action \cite{tot09,tot18}. 
In any case, this not only resulted in a new concept (e.g., PERCUSSION) but also modified the affordances (in the sense of \cite{gib}) of STONE (i.e., stone was now something that could be fractured through percussive action).

\begin{figure}
\centering
\includegraphics[scale=0.45]{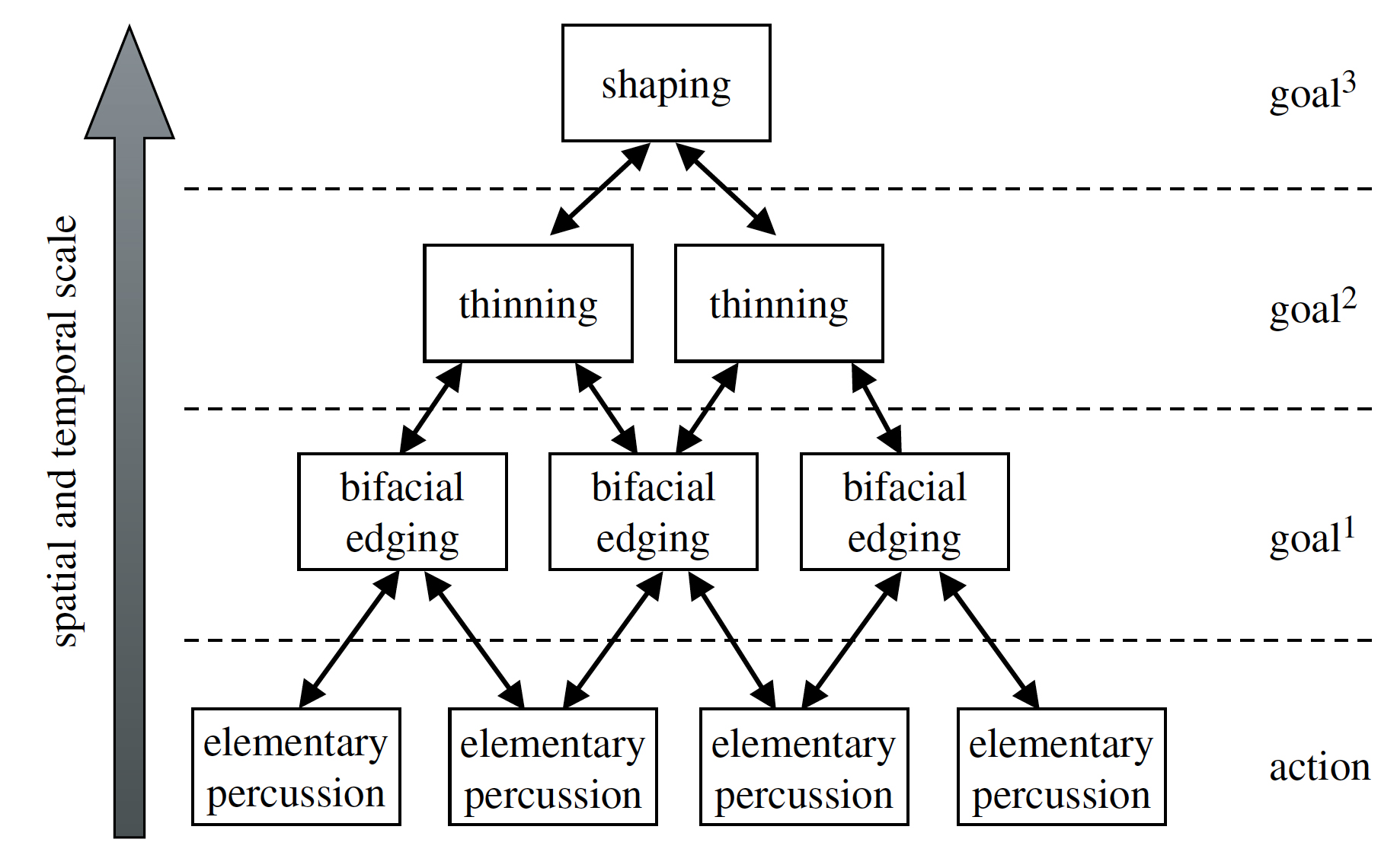}
\caption{Hierarchical organization of tasks in Acheulean toolmaking. From Stout, 2008.}
\label{fig-subgoals}
\end{figure}

Although there may have been some degree of convergent evolution, Oldowan technology was transmitted through {\it social learning} processes such as imitation or guided instruction, from one generation to the next, and Acheulean technology built on this pre-existing technology, as indicated in Fig.~\ref{fig-subgoals}.  
To make an Acheulean hand axe required (1) skillful coordination of perception and motor skills involving recursive modification of an action according to the outcome of the previous action, and (2) shifting between hierarchically organized (short-term) sub-goals at different spatiotemporal scales to achieve the desired (long-term) outcome \cite{sto08,ini99}.
The toolmaker had to bear in mind not just the current and desired final states of the tool but also the multiple procedural actions---edging, thinning, and shaping---required to achieve the final state. Since these actions were not yet observable in the environment, the concepts EDGING, THINNING, and SHAPING had to be generated from scratch.
  
Given the cumulative increase in the sophistication of the Acheulean hand axe over time (see Fig.~\ref{oldowan-vs-Acheulean}), it appears not to have been the handiwork of one individual, but rather invented collectively, with the intentional scaffolding of each new improvement (e.g., edging, thinning, and shaping), potentially separated by generations. 
Note that not only do EDGING, THINNING, and SHAPING constitute new MRs but, collectively, they constitute a cMR, that of how to make an Acheulean hand axe. 
Moreover, each new concept further modified the perceived affordances of STONE (i.e., stone was no longer just something that could be fractured with percussive action, but something that could also be edged).
Although (as mentioned previously) it is generally believed that complex language was not yet possible, social learning processes involving demonstration and/or imitation of the finished product would have enabled spatiotemporal diffusion of these `partial solutions.'\footnote{Non-verbal transmission of stone-tool technologies and its relationship to language evolution is discussed further in \cite{morgan15,ohnuma97}.}

\section{RAF model of the invention of the Acheulean hand axe}

We now introduce the mathematical framework and terminology that will be used to model the invention of the Oldowan and Acheulean tool technologies.

\subsection{\bf Terminology}

All mental representations (MRs) in a given individual $i$ are denoted $X_i$, and a particular MR $x_i$ in $X_i$ is denoted by writing $x \in X_i$.
As in an OOL RAF, we have a {\it food set};  for individual $i$, denoted $F_i$. In the OOC context, $F_i$ encompasses MRs for individual $i$ that are either innate, or that result from direct experience in the world, including natural, artificial, and social stimuli. 
$F_i$ includes everything in the long-term memory of individual $i$ that was not the direct result of individual $i$ engaging in RR. 
This includes information obtained through social learning from {\it someone else} who may have obtained it by way of RR. For example, if individual $i$ learns from individual $j$ how to edge a blank flake through percussive action, this is an instance of social learning, and the concept EDGING is therefore a member of $F_i$.

$F_i$ also includes existing information obtained by $i$ through individual learning (which, as stated earlier, involves learning from the environment by nonsocial means), so long as this information retains the form in which it was originally perceived (and does not undergo redescription or restructuring through abstract thought).
The crucial distinction between food set and non-food set items is not whether another person was involved, nor whether the MR was originally obtained through abstract thought (by {\it someone}), but whether the abstract thought process originated in the mind of the individual $i$ in question. 
Thus, $F_i$ has two components:
\smallskip
 \begin{enumerate} [leftmargin=*,labelindent=15pt,label= \arabic*.]
\item  ${\mathbb S}_i$ denotes the set of MRs arising through direct stimulus experience that have been encoded in individual $i$'s memory. It includes MRs obtained through social learning from the communication of an MR $x_j$ by another individual $j$, denoted ${\mathbb S}_i[x_j]$, and MRs obtained through individual learning, denoted ${\mathbb S}_i[l]$, as well as contents of memory arising through direct perception that do not involve learning, denoted ${\mathbb S}_i[p]$.
\smallskip
\item $I_i$ denotes any {\em innate knowledge} with which individual $i$ is born.
\end{enumerate}
\smallskip

A particular catalytic event (i.e., a single instance of RR) in a stream of abstract thought in individual $i$ is referred to as a {\it reaction,} and denoted $r \in \cR_i$. A stream of abstract thought, involving the generation of representations that go beyond what has been directly observed, is modeled as a sequence of catalytic events. Following \cite{gabste}, we refer to this as a {\it cognitive catalytic process} (CCP). 
The set of reactions that can be catalyzed by a given MR $x$ in individual $i$ is denoted $C_i[x]$. 
The entire set of MRs either {\it undergoing} or {\it resulting from} $r$ is written $A$ or $B$, respectively, and a member of the set of MRs undergoing or resulting from reaction $r$ is denoted $a \in A$ or $b \in B$. 
     
The term {\it food set derived}, denoted $\neg F_i$, refers to mental contents that are {\it not} part of $F_i$, i.e., the products of any reactions derived from $F_i$ and encoded in individual $i$'s memory. Its contents come about through mental operations {\it by the individual in question} on the food set; in other words, food set derived items are the direct product of RR. Thus, $\neg F_i$ includes everything in long-term memory that {\it was} the result of one's own CCPs. 
It may include a MR in which social learning played a role, so long as the most recent modification to this MR was a catalytic event ( i.e., it involved RR).\footnote{This distinction between food set and food set derived may not be so black and white but for simplicity we avoid that subtlety for now.} $\neg F_i$ consists of all the products $b \in B$ of all reactions $r \in R_i$.

The set of {\it all} possible reactions in individual $i$ is denoted $\cR_i$.
The mental contents of the mind, including all MRs and all RR events is denoted $X_i \oplus \cR_i$. This includes $F_i$ and $\neg F_i$. 
Recall that the set of all MRs in individual $i$, including both the food set and elements derived from that food set, is denoted $X_i$. 
$\cR_i$ and $C_i$ are not prescribed in advance; because $C_i$ includes remindings and associations on the basis of one or more shared property, different CCPs can occur through interactions amongst MRs. Nevertheless, it makes perfect mathematical sense to talk about $\cR_i$ and $C_i$ as sets. 
Table~\ref{OOL-OOC} summarizes the terminology and correspondences between the OOL and the OOC.

\begin{table}[ht]
\caption{Terminology and correspondences between the Origin of Life (OOL) and the Origin of Culture (OOC).}
\begin{center}
\begin{tabular}{@{} llllllllllllll @{}}
\hline \hline 
 \textbf{Term} & \textbf{Origin of Life (OOL)} & \textbf{Origin of Culture (OOC)} \\ 
    \hline
     $X_i$ & all molecule types in protocell $i$ & all mental representations (MRs) in individual $i$\\
     $x \in X_i$ & a molecule in $X_i$  & a MR in $X_i$ \\
     $F_i$ & food set for protocell $i$  & innate or directly experienced MRs by $i$ \\
     $r \in \cR_i$ & a particular reaction in $i$ & a particular representational redescription (RR) in $i$\\
     $C_i[x]$ & reactions catalyzed by $x$ in $i$ & RR events `catalyzed' by $x$ in $i$\\
     $(x,r) \in C$ & $x$ catalyzes $r$ & $x$ `catalyzes' redescription of $r$ \\
     $a \in A$ & member of set of reactants in $r$ & member of set of MRs undergoing $r$\\ 
     $b \in B$ & member of set of products of $r$ & member of set of MRs resulting from $r$ \\ 
     $\neg F_i$ & non food set for $i$; (i.e., all $B$ of $\cR_i$) & MRs resulting from $R_i$;  (i.e., all $B$ of $R_i$) \\
\hline
\hline
\end{tabular}
\end{center}
\label{OOL-OOC}
\end{table}

Our model includes elements of cognition that have no obvious parallel in the OOL. We denoted the subject of attention at time $t$ as $\ring{w_t}$. It may be an external stimulus, or a MR retrieved from memory.
Any other contents of $X_i \oplus \cR_i$ that are accessible to working memory, such as close associates of $\ring{w_t}$, or recently attended MRs, were denoted $W_t$, with $W_t$ constituting a very small subset of $X_i \oplus \cR_i$.
In the present paper, the focus is not on these OOC-specific components of the model in order to tackle the question of how non-food set derived MRs (i.e., a non-empty $\neg F$) came about, because these non-food set derived MRs are essential to the emergence of a semantic network that is self-organizing and autocatalytic. 

\begin{figure}
\centering
\includegraphics[scale=0.21]{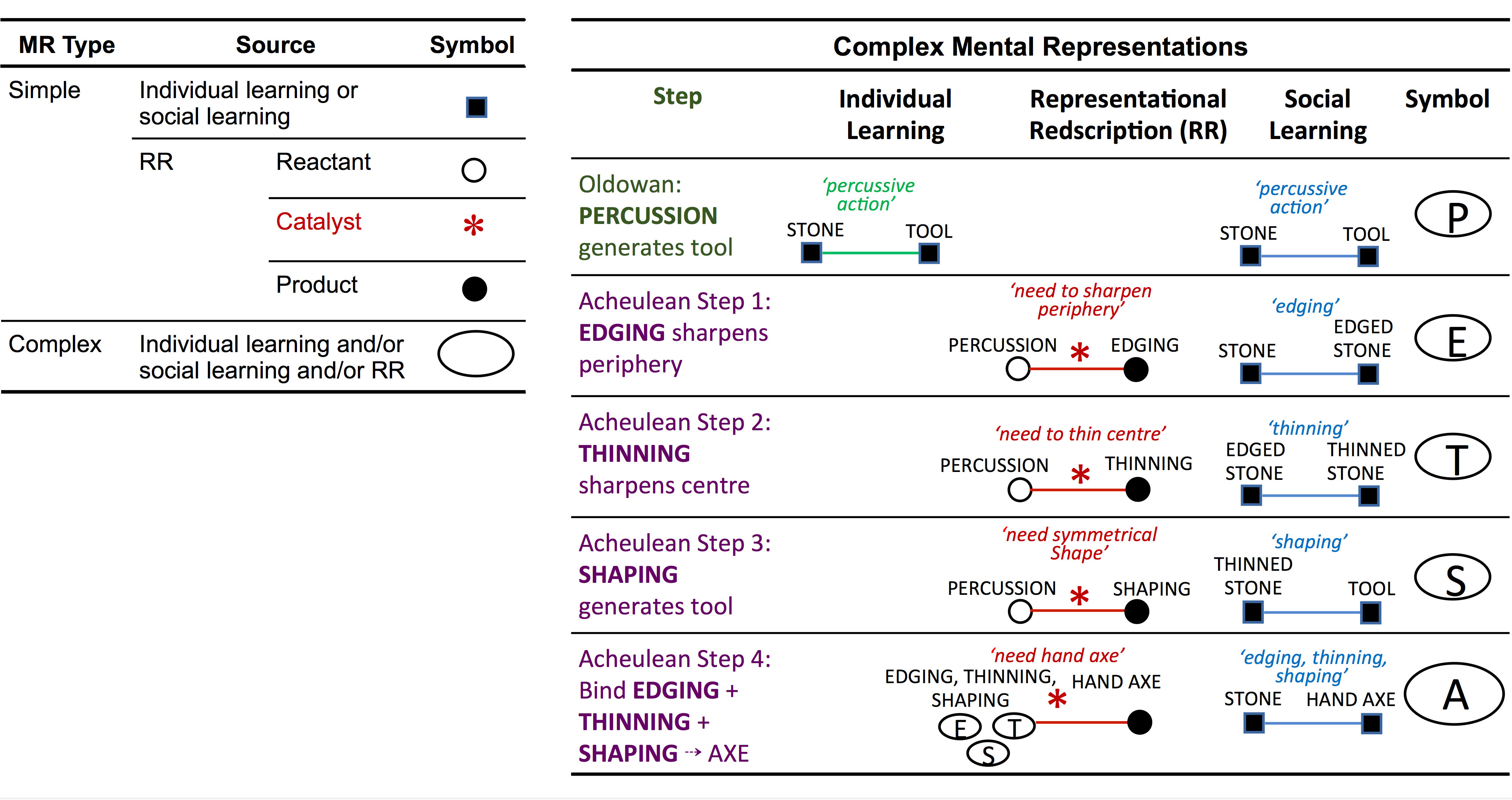}
\caption{Left: Sources of simple and complex mental representations (sMRs and cMRs) and symbols used here to depict them. Right: cMRs and the sMRs  of which they are composed, involved in the invention of the Oldowan tool (top row) and the Acheulean hand axe (lower rows). For each tool, only one scenario discussed in the text is portrayed here. Each instance of social learning (Column 4) must be preceded by a relevant instance of individual learning (Column 2) or representational redescription (RR) / abstract thought (Column 3).}
\label{steps}
\end{figure}

\begin{figure}
\centering
\includegraphics[scale=0.25]{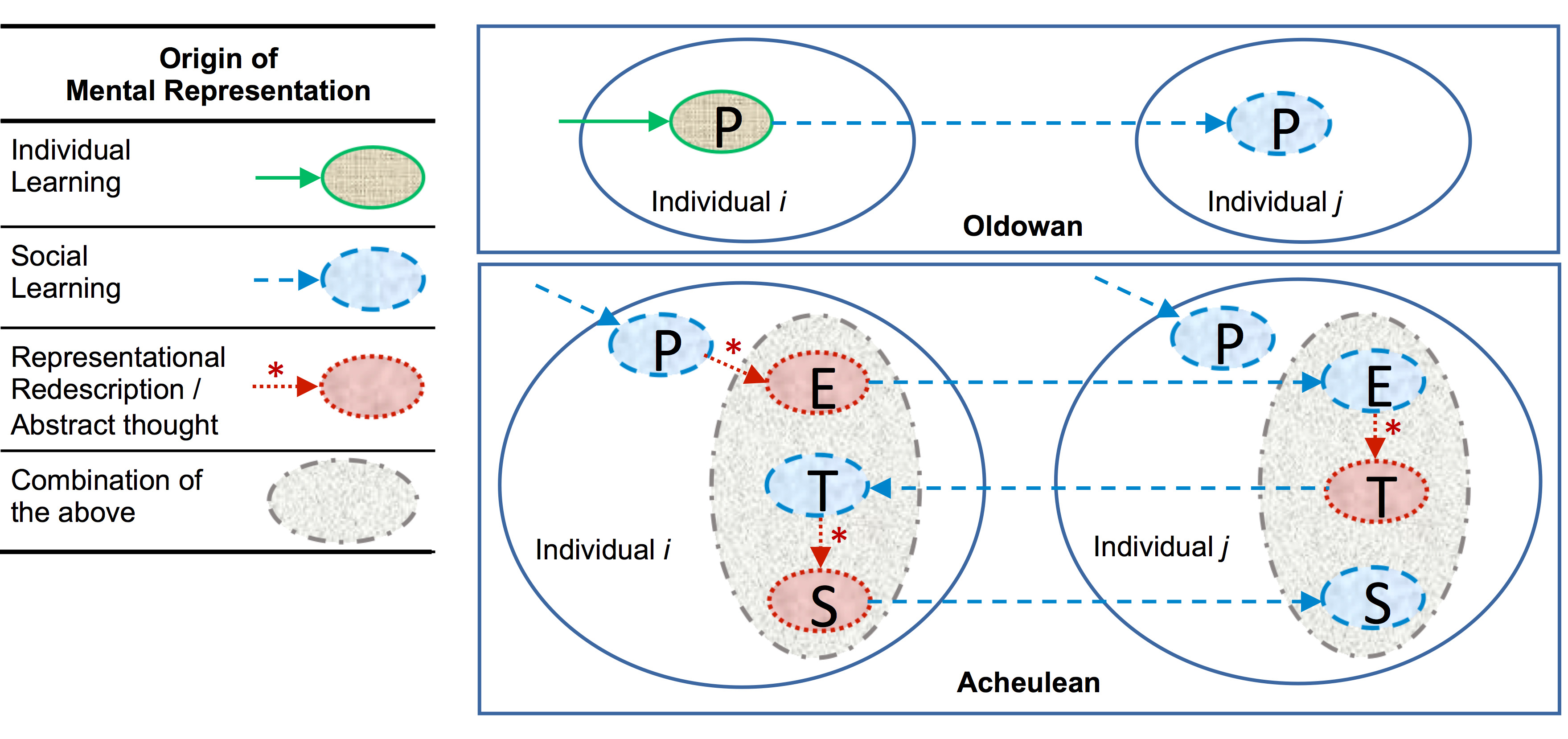}
\caption{Left: Key showing the depiction of complex MRs arising through individual learning, social learning, representational redescription (RR) or abstract thought, or a cominbation of these.  Right: Schematic representation of possible events culminating in the invention of Oldowan (top) and Acheulean (bottom) tools. Top right: Individual $i$ obtains an understanding of percussion through individual learning by watching falling rocks splinter the stone below. Individual $i$ imitates the action of the falling rock on a piece of stone, thereby creating a tool. Individual $j$ obtains the toolmaking technique from individual $i$ through social learning. Bottom right: Individuals $i$ and $j$ both acquire the concept of PERCUSSION through social learning from their parents, but only in $i$ does it undergo catalysis to generate the concept of EDGING, which $i$ shares with $j$ through social learning. Similarly, $j$ invents the concept THINNING and shares it with $i$, and $i$ invents the concept SHAPING and shares it with $j$. At this point, they both possess the entire skill set to make an Acheulean hand axe. Note that although for simplicity this sequence is portrayed here with only two individuals, in reality the multiple stages in the invention of this tool likely involved numerous individuals spanning generations.
As in Fig.~\ref{steps}, only one of the scenarios discussed in the text is portrayed here.}
\label{fig:origin}
\end{figure}

Now that the mathematical framework has been introduced, let us compare the cognitive processes involved in the invention of Oldowan versus Acheulean tools from the perspective of RAFs using an autocatalytic framework. This is illustrated in Fig.~\ref{steps} and Fig.~\ref{fig:origin}. 

\subsection{\bf Oldowan}

We begin by modeling the invention of an Oldowan tool by individual learning and imitation of the possible effect of percussive action on stone, arising through observation of accidental breakage (as discussed above). The invention involves noticing that percussive action increases the capacity of a rock to be used as a tool. It involves the formation of an association between the concepts ROCK and TOOL, and results in a cMR that relates stone, percussive action, and tool.

Note that the inventor's thought process involves recursion, in the sense that the output of a previous percussive action serves as the input to the decision of whether to continue, and recursive processing continues until the task is complete. However, since no non food set MRs are derived, there is no `internally catalyzed reaction' as we are defining it. Therefore, from a RAF perspective, the mind of an early hominid that relied on Oldowan technology can be described as one for which all mental contents are members of the food set of innate or directly experienced MRs. Thus, the set of non food set MRs is empty. In terms of the formalism we are using, $\cQ = (X, F)$; in other words, we need not consider $\cR$ and $C$.

For individual $k$ ($k= i,j$), let $P_k$ be the complex MR that combines the sMRs STONE, PERCUSSION, and TOOL. Thus, in the Oldowan setting illustrated in Fig.~\ref{fig:origin}, the generation of a tool through either individual learning (in $i$) or social learning (in $j$) amounts to adding $P_k$ to the food set of individual $k$.  Formally, we can write this as:
$$F_i \mapsto F_i \cup \{P_i\}, \mbox{ where } P_i \in {\mathbb S}_i(l),$$
and
$$F_j \mapsto F_j \cup \{P_j\}, \mbox{ where } P_j \in {\mathbb S}_j(P_i).$$
Thus, the first cMR results from individual learning, whereas the second arises from social learning by individual $j$ of the concept $P_i$ from individual $i$.

Note that although the sMRs STONE, PERCUSSION, and TOOL are connected by associations, these associations were not obtained through abstract thought, but through observations of cause and effect in the external world. Since invention of the Oldowan tool required only one cMR (the CMR of how to make an Oldowan tool), there would have been no back-and-forth social exchange of partial solutions involved in its invention, and there are no higher-level cMRs.
 
\subsection{\bf Acheulean}

Acheulean tools were the culmination of several (perhaps spatiotemporally separated) steps that intertwined individual learning, social learning, and creative thinking, as illustrated in Fig.~\ref{fig:origin}. To model the cultural transition from Oldowan to Acheulean technology, we begin with social transmission of the cMR for the process of making an Oldowan tool. In the mind of each individual that acquired this cMR through social learning, the food set was enlarged, as described above. Social transmission of this cMR continued for generations before it was elaborated. 
 
Elaboration entailed three insights into new ways of processing the basic Oldwowan tool to make it more useful. These insights are represented as `catalysis events' because they resulted in the generation of something new: a new MR. Each catalysis event came about through a CCP and resulted in a new non food set item. Catalyzed reactions transform an element of $F_k$ into an element of $\neg F_k$. 
For an individual $k$ (where $k=i,j$), we write  
$a_k \xrightarrow{b_k} c_k$ to denote the reaction that transforms one MR ($a_k$) to a resulting MR $c_k$ (in $\neg F_k$) by catalyst $b_k$.

The first catalysis event was provoked by the context: {\it need to sharpen periphery}. This context was internally represented as a thought, or MR. It caused modification of the {\it reactant}, PERCUSSION, to generate a {\it product}: the new concept EDGING. Since EDGING is connected through association to ROCK, this event expands the affordances of ROCK (i.e., it is now perceived as something that can be edged). 
Affordances are a kind of association and, as such, they increase the connectivity of the conceptual network. 
We represent this by a catalyzed reaction (and the associated individual learning) as follows: 

 \begin{equation}
 \label{eieq}
     P_i \xrightarrow{p_i} E_i,  \mbox{ } F_i \mapsto F_i \cup \{E_i\},
     \end{equation}
 where $p_i$ is the catalyst MR {\it need to sharpen periphery}, and $E_i \in {\mathbb S}(l)$ is the resulting cMR.

The specifics of the situation that inspired the invention---what we are modeling as the context that `catalyzed' this event---need not be socially transmitted. Thus, what was invented is not necessarily identical to what is socially transmitted. It is the new affordance of ROCK and its association with the new concept of EDGING that are socially transmitted. 
We represent this transmission event as follows:
\begin{equation}
\label{eiieq}
F_j \mapsto F_j \cup \{E_j\}, \mbox{ }  E_j \in {\mathbb S}_j(E_i).
\end{equation}
{\indent}In the second catalysis event, the context was {\it need to thin the center}. The reactant EDGING was modified to generate another product, the new concept of THINNING.
This can be represented as follows:
\begin{equation}
\label{eiiieq}
E_j \xrightarrow{t_j}  T_j,  \mbox{ } F_j \mapsto F_j \cup \{T_j\},\end{equation}
where $t_i$ is the catalyst MR `need to thin centre' and $T_j \in {\mathbb S}_j(l)$ is the resulting cMR. 
Similarly, in the context, {\it need to create symmetrical shape}, the reactant---the concept THINNING---was modified to generate another product, the concept of SHAPING. 
The formal description of these two catalytic processes is analogous to that of the formation of the EDGING cMR.

To summarize, the cultural evolution of Acheulean technology depicted in Fig.~\ref{fig:origin} is described formally in the following sequence of six processes, where steps (i) to (iii) correspond to equations  (\ref{eieq}) to (\ref{eiiieq}) above. Again, although for simplicity we consider only two individuals, the steps described as catalyzed reaction were likely contributed by individuals separated across generations. Catalyzed reactions take place in steps (i), (iii) and (v). For example, in step (v), the catalyst is the MR of `need symmetrical shape', denoted $s_i$. $S_i$ and $S_j$ are the resulting complex MRs in individuals $i$ and $j$, respectively.

\vspace{\baselineskip}
\indent
     \parbox{\textwidth}{%
 \begin{itemize}
\item[(i)] [RR leading to $E$ by $i$]  $P_i \xrightarrow{p_i} E_i, F_i \mapsto F_i \cup \{E_i\};$
\item[(ii)] [social learning of $E$ from $i$ by $j$] $F_j \mapsto F_j \cup \{E_j\}, E_j \in {\mathbb S}_j(E_i);$
\item[(iii)] [RR leading to $T$ by $j$] $E_j \xrightarrow{t_j} T_j, F_j \mapsto F_j \cup \{T_j\}$;
\item[(iv)] [social learning of $T$ from $j$ by $i$] $F_i \mapsto F_i \cup \{T_i\}, T_i \in {\mathbb S}_i(T_j)$;
\item[(v)] [RR leading to $S$ by $i$] $T_i \xrightarrow{s_i} S_i, F_i \mapsto F_i \cup \{S_i\}$;
\item[(vi)] [social learning of $S$ from $i$ by $j$] $F_j \mapsto F_j \cup \{S_j\}, S_j \in {\mathbb S}_j (S_i).$
\end{itemize}
}

\vspace{\baselineskip}

The formation of abstract concepts such as SHAPING was essential for the emergence of more extensive autocatalytic structure because abstract concepts tend to become more densely connected through associations than superficial concepts. As extreme examples, the concepts DEPTH and OPPOSITE are relevant to almost every knowledge domain. The concepts that arose in the invention of the hand axe may be less widely applicable than concepts such as OPPOSITE, but they could potentially be applied to other domains (such as food preparation). Abstract concepts create new affordances for existing concepts (e.g., the concept SHAPING could have create affordances for ANIMAL SKIN that enabled it to be turned into clothing). This then further increased the density of associations.

\begin{figure}
\centering
\includegraphics[scale=0.55]{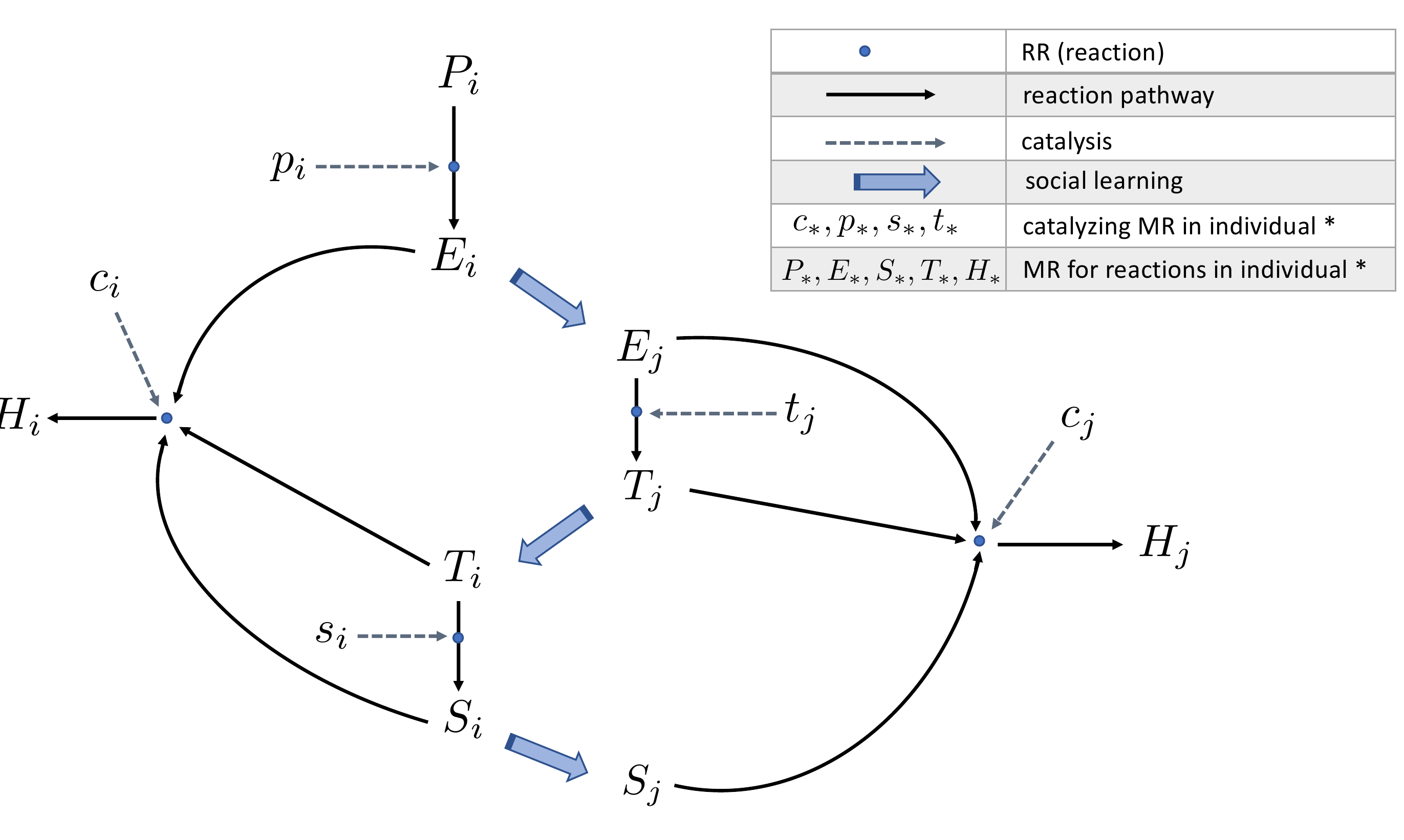}
\caption{Part of the RAF structure involved in the invention of the Acheulean hand axe.
Again, although for simplicity we have only individuals $i$ and $j$, these steps were likely spread out spatiotemporally. The figure depicts the processes labeled (i)--(vi) in the text, as well as the process of combining the three steps EDGING, THINNING, and SHAPING, into the final mental representation in each of the two individuals of how to make an Acheulean hand axe, $Hi$ and $Hj$.
}
\label{raf}
\end{figure}

The Acheulean hand axe required not just the invention of the three cMRs associated with each phase of the toolmaking process, but also recursive RR on them so as to generate an even more complex MR: the representation of the entire process of making a hand axe. 
If we denote this complex meta-RR (capturing the entire process of making a hand axe) in  individual $i$ by $H_i$, then we can describe the generation of $H_i$ by the (catalysed) reaction:
\begin{equation}
    \label{threeq}
    E_i + T_i+S_i \xrightarrow{c_i} H_i,
\end{equation}
where the catalyst $c_i$ is the realization by individual $i$ that combining $E$, $S$, and $T$ will lead to the desired tool.\footnote{Although successful completion of a task may result in neural signals that reinforce the eliciting behavior, that happens after the action has taken place; it is not the catalyst. The catalyst acts before the action has taken place. It is the thought that prompts the action to take place.}
The recognition (on the part of both individual $i$ and individual $j$) that the entire series of steps can be clumped together as `how to make an effective tool' constitutes yet another catalyzed reaction, as indicated in Fig.~\ref{raf}.

The invention of the Acheulean hand axe corresponds to the emergence of RAF sets of MRs that are close associates, and accessible to one another. In particular, RR enables the content of working memory to be updated through abstract thought or reflection drawing on content from long-term memory,\footnote{Creative insights (i.e., those that make significant contributions to culture) often arise subconsciously; i.e., they arise from just beyond the confines of working memory \cite{bowetal}.} with or without an environmental stimulus acting as a reminder or cue. We describe this more formally as follows, following \cite{gabste}.

We denote each MR in a RR reflection process as $m \in M_t$. We say that $\ring{w} \in W_t$ is catalyzed by an item $m \in M_t$. This `reaction' updates the subject of thought, which is now denoted $\ring{w'} \in W_{t+\delta}$. A single step RR (referred to in \cite{gabste} as {\em cognitive updating}) is denoted:

 $$\ring{w} \xrightarrow{m} \ring{w'}, \mbox{ and } \ring{w} \mapsto w.$$

A sequence of recursive RR events, which, as mentioned above, is referred to as a CCP, is described as: $$\cC= \ring{w}_{t(1)}, \ring{w}_{t(2)}, \ldots, \ring{w}_{t(k)},$$ (where $\ring{w}_{t(i)} \in W_{t(i)}$ and where $t(i)$ values are increasing), such that each MR $\ring{w}_{t(i)}$ is the reactant catalyzed by an environmental stimulus or MR from memory to generate a new MR, the product of that reaction.

Thus, the CCP that connected EDGING, THINNING, and SHAPING resulted in a meta-cMR composed of three hierarchically structured cMRs, which were themselves composed of simple MRs. This constituted an important step toward a self-organizing, autocatalytic structure, not just because the CCPs forged associations between items in memory but because (as discussed above) the abstract concepts EDGING, THINNING, and SHAPING could potentially be applied to other domains, creating still more associations. 

\subsection{\bf Cognitive RAFs}

We now ask: does the Acheulean mind as described above contain a genuine RAF? To answer this question, we must consider the mean rate at which RR reactions are taking place, denoted $\lambda$, and whether or not this exceeds a certain threshold. We can describe this more precisely as follows (from \cite{gabste}, where a proof is provided).\footnote{It is assumed that $t$ is sufficiently large that RR reactions have commenced, and that the rate at which new environmental stimuli appear is bounded.}

\begin{proposition}
\label{mainpro}
$\cQ$ contains a RAF that increases in size with time $t$ (namely the set of RR reactions that actually occur between time 0 and $t$). Moreover, while $\lambda$ is below a critical threshold, CCPs in this RAF are short and few in number, but when  $\lambda$  exceeds this threshold, CCPs become more frequent, persistent and complex.
\end{proposition}

The RAF described in Proposition~\ref{mainpro} has the additional feature of being a `constructively autocatalytic F-generated set', as defined in \cite{mos}. Such a RAF has the property of being self-organising, and able to self-replicate and evolve (albeit in an inefficient manner, without a self-assembly code). 
In the current cultural context, we refer to such a RAF as a persistent cognitive RAF, or simply, a {\em cognitive RAF}. In our cultural context, $\lambda$ is expected to vary positively with the complexity of the existing network structure (more ways to reflect on something new) and negatively with the degree to which one's internal model already appears to faithfully capture the content of one's environment (nothing left to think through). 

Returning to the question with which we began this section, although the mind of the Acheulean toolmaker did not achieve a cognitive RAF, it achieved what could be called a {\it transient RAF} because it contains the cognitive equivalent of catalyzed reactions (as discussed above), and it appears, for an instant, like a RAF. A transient RAF is not self-organizing, and therefore it cannot generate open-ended cultural novelty. 
However, compared to the mind of the Oldowan toolmaker, in which (as far as we know from the archaeological evidence) there were no catalyzed reactions, it marks the crossing of a significant hurdle toward the achievement of genuine, persistent RAF structure.

The next question is: once hominids were capable of recursive, hierarchically structured thought and creative problem solving, why was the invention of the Acheulean hand axe followed by approximately one million years of cultural stasis \cite{tat98}?
This is an open and much-pondered question in the archaeological literature, and the autocatalytic approach to culture developed here suggests a tentative answer. Specifically, our model suggests that this was because $\lambda$ (the mean rate at which RR reactions occur) did not rise above a critical threshold to generate self-sustained cognitive reorganization, meaning that any CCPs that arose were short and few in number.

\subsection{\bf Self-replication of transient cognitive RAF}

Although in the mind of the Acheulean toolmaker the RAF was only transient, and did not self-organize and evolve in an open-ended manner, it yielded an ongoing dynamical process nonetheless, by way of its influence on other individuals. As mentioned earlier, based on current thinking, the above model assumes that although early hominids that did not yet possess complex language, the results of creative thought---modelled here as CCPs---could extend beyond a single individual through social learning and pedagogy \cite{teh08}. This means that a social group can be described as a higher-level transient RAF that follows the same processes as described earlier. 

We model this as follows. Given a social group $\G$ composed of individuals $i$, $j$, ..., the collection $M$ of all MRs in the social group is the disjoint union of the MRs in each individual mind. The union is disjoint because $m_i$ and $m_j$ refer to MRs in different individuals ($i$ and $j$). 
When the MRs $m_i$ and $m_j$ in two individuals concern the same feature of the world, we say that $m_i$ and $m_j$ are {\em homologous}, denoted by writing  $m_i \sim m_j$.
If individual $j$ provides the context that triggers a RR event or creative insight in another individual $i$, the process of cognitive change extends across two individuals. More precisely, if a reaction $\ring{w}_i \xrightarrow{s} \ring{w'}_i$ in individual $i$ is catalyzed by a MR $s \in {\mathbb S}(w_j)$ then we can (formally) regard this as the catalyzed reaction
\begin{equation}
\ring{w}_i \xrightarrow{w_j} \ring{w'}_i.
\label{eqo}
\end{equation}

Note that the MRs involved in this reaction as  reactant and product are in the  mind of one individual ($i$), whereas the catalyst is in the mind of another ($j$). Therefore, we obtain  an equivalence ($\equiv$) between  social learning  and a cognitive reaction (involving a catalyst in a different mind). This leads to the notion of a RAF that consists of reactions and CCPs within and between the items in the collective minds of the social group. 
Social learning processes enable a MR (such as the procedure used to produce a hand axe) to become established as homologous MRs across individuals, and spread amongst group members in $\G$, as illustrated in Fig.~\ref{raf}. Thus, the transient RAF replicates across individuals of a social group. 
The size of the resulting CCPs in the group as a whole is influenced by two parameters.  The first is the structure of the digraph (directed graph) $\D$ with vertex set $\G$ and an arc from $i$ to $j$ if individual $i$ is able to communicate concepts to individual $j$. Note that $\D$ depends on time; for example, new groups of individuals form and split over time.  

The connectedness of $\D$ could vary from a single individual with arcs to and from all others (a political leader, celebrity, or `guru'), to a network in which each individual has an arc to every other individual.
A second parameter, denoted $\rho$, scales the rate at which catalysis events of the type described in Reaction (\ref{eqo}) above occur when $(j,i)$ is an arc of $\D$.  Thus, when $\rho=0$, it is never the case that one individual provides the context that triggers RR in another.
In reality, the rate of catalysis for the arc $(j,i)$ may depend more finely on $i$ and $j$, so $\rho$ is treated as an overall scaling factor. The response of CCPs in the social group to increasing $\rho$ (presumably related to the emergence of increasingly sophisticated communication) is summarized in the following result (an analogue of Proposition 1 but at the level of the social group). Recall that any  digraph has a unique composition into strongly connected components.  The following result follows from a simple percolation argument on directed graphs (see Appendix). 

\begin{proposition}
For small values of $\rho$, most CCPs occur within individuals, and homologous items tend to occur only between closely linked individuals (in $\D$). As $\rho$ increases, CCPs grow in size and involve longer chains of catalyzation events, leading to homologous items spreading throughout each strongly connected component of $\D$.
\end{proposition}

\section{Discussion and conclusions}

To understand how cultural evolution got started, we must ask what kind of semantic structure would be capable of initiating and sustaining open-ended cultural change, and examine how such structure came about in the minds of our ancestors. Building on earlier work \cite{gab98,gabste,glvl11,vel12}, 
this paper examines early {\em Homo} cognition through a particular lens: the emergence of semantic structure that is self-organizing, self-reproducing, and autocatalytic. Of course, minds are part of living organisms, which have these properties, but we are interested in the emergence of a second-order level of self-organizing structure that pertains not to cellular or organismal processes but to the webs of associations by which hominids weave together an understanding of their world. We suspect that these properties were as important to cultural evolution as they were to biological evolution. Since autocatalytic networks possess these properties and have been useful in modeling the origin of life, we used an autocatalytic RAF framework to model what is arguably the earliest significant transition in the archaeological record: the transition from Oldowan to Acheulean tool technology. 

We hope that future research will build on this direction by comparing the cultural RAF approach developed here with other standard semantic network approaches \cite{bea16,bar13}. Although these standard semantic networks suffice for modeling semantic structure in individuals, we believe that the RAF approach will turn out to be superior for modeling lineages of cumulative cultural change, because it distinguishes semantic structure arising through social or individual learning (modeled as food set items) from semantic structure {\it derived from} this pre-existing material (modeled as non food set items generated through abstract thought processes that play the role of catalyzed reactions). This makes it feasible to model how cognitive structure emerges, and to trace lineages of cumulative cultural change step by step. It also frames this project within the overarching scientific enterprise of understanding how evolutionary processes (be they biological or cultural) begin, and unfold over time. Data for comparing cultural RAFs against other semantic networks could come not just from existing archaeological data sets and analyses of social learning and pedagogy in stone toolmaking \cite{teh08}, but from neuroscience, building on neuroscientific studies of brain activation during toolmaking by modern-day novice and expert Oldowan and Acheulean toolmakers \cite{sto08}. We note that there now exist established methods for developing semantic networks using neuroscientific data \cite{bet17,kar16,med15}. To our knowledge these methods have not yet been used in the study of cultural lineages but we see no reason why they couldn't be.

The model as it stands has limitations; it is highly simplified, and we do not precisely know the context in which the cognitive events modelled here took place.
Future versions could incorporate a more sophisticated representation of interactions amongst MRs \cite{aeretal16, aeretal13} and a dynamic representation of context \cite{howkah02,vel11}. There is also more work to be done on the implications of cognitive RAF theory for the evolution of cooperation (see \cite{vor20}), and the evolutionary pressures shaping cognitive RAFs (a promising effort in this direction is \cite{andtor19}).

Since there are multiple ways that a given set of MRs can be networked together into a viable cognitive RAF, this model may also provide a new approach to understanding the origins of psychological differences at the individual \cite{sac17} and cultural \cite{sngetal18} levels.
The question arises as to how the inventors of new concepts (such as EDGING) differed from their less creative kin may be due to individual differences in two parameters of our model: (1) reactivity (the extent to which the meaning of a MR is perceived to be altered in an interaction---or `reaction'---with another MR), and (2) $\lambda$ (the rate at which reactions occur). For example, what is referred to as cognitive rigidity may be a matter of low $\lambda$, resulting in a semantic network that is {\it subcritical}, whereas creative thinking may be a matter of high $\lambda$, resulting in a semantic network that is {\it supracritical}. Which of these two regimes a particular individual falls into may also depend on variables associated with the `food set' MRs, including the degree of detail in which these MRs were encoded, and their diversity. For example, the food set MRs will be more diverse for individuals that have experienced different climates, environments, or cultures.

Although the invention of the Acheulean hand axe modeled here resulted in genuine albeit transient autocatalytic structure, as mentioned earlier, archaeological evidence suggests that over a million years passed before the emergence of a persistent cognitive RAF. 
In another paper (in progress), we propose that  rapid cultural change in the Middle-Upper Paleolithic between 100,000 and 50,000 years ago required the ability to, not just recursively redescribe the contents of thought, but reflect on ideas from widely varying perspectives, at different levels of abstraction, and that this in turn required the capacity to tailor the reactivity of thought to the current situation.
This resulted in a second cognitive transition culminating in the crossing of a percolation threshold yielding a self-organizing autocatalytic semantic network that extended across different knowledge domains, and routinely integrated new information by reframing it in terms of current understandings.  
The proposal is consistent with there being a genetic basis to cognitive modernity \cite{cor04}, except that onset of the capacity for variable reactivity would have underwritten, not just complex language, but also other cultural innovations of this period, such as the ability to adapt tools to widely differing task-specific uses, and generate art that served utilitarian, decorative, and possibly religious purposes. Thus, the model developed here can be extended beyond the Oldowan-Acheulean transition and applied to other periods of cultural change such as the Middle-Upper Paleolithic, and potentially also, the explosion of cultural novelty we are witnessing currently. 

There are formal models of many aspects of human cognition, such as learning, memory, planning, and concept combination, but little in the way of formal models of how they came to function together as an integrated whole, and how such wholes affect one another over the course of human history.
The structure of a modern human mind serves as a scaffold for the interpretation of both external and internally generated MRs, which perpetually reinforce and revise that structure. It generates a unique stream of thought and experience, which expresses itself through its contribution---through a smile, a turn of phrase, or a world-changing innovation---to cultural evolution. 
The RAF approach taken here provides a means of addressing how this kind of semantic network came about, and how it evolves over time. 

\newpage

\bibliographystyle{apacite}
\bibliography{cogsci}
\newpage
\section{Appendix}

{\em Proof of Proposition 2:}
If we treat the transfer of CCPs between individuals (under the process described by Expression (\ref{eqo})) as a stationary Markov process on the directed graph $\D=(V,A)$, the probability $P$ that no such events occur in the time interval $[0, T]$ is given by:
\begin{equation}
\label{pro1}
P= \prod_{v \in V} [\exp(-c\rho T)]^{o(v)} = \exp(-c\rho T |A|),
\end{equation}
 where $o(v)$ is the out-degree of $v$, and $c>0$ is a constant (note that $\sum_{v \in V} o(v) = |A|$). It follows from Eqn.~ (\ref{pro1}) that $P$ converges to 1 as $\rho \rightarrow 0$.  
 
 On the other hand, suppose that  $v$ lies in a strongly connected component $C$ of $\D$ and let $d$ denote the smallest integer for which, for every vertex $w$ in $C$, there is a directed path from $v$ to $w$ of length at most $d$ ($d$ is well-defined, since $v$ and $w$ lie in the same strongly connected component of $\D$).  Then, for any other individual $w$ in $C$, the probability that at least one CCP percolates from $v$ to $w$ in the time interval $[0,T]$ (either directly or via a chain of other individuals in $C$) is at least:
\begin{equation}
\label{pro2}
\left(1-\exp\left(-c'\rho\frac{T}{d}\right)\right)^d,
\end{equation} 
and this quantity clearly converges to 1 as $\rho$ grows (here, $c'>0$ is a constant). 
To justify the claim that Expression~(\ref{pro2}) is indeed a lower bound on the stated probability, observe that if $v = v_0, v_1, v_2, \ldots, v_k$ is any path in $C$ of length $k\leq d$, then the probability that a given CCP percolates from $v_i$ to $v_{i+1}$ (where $0 \leq i < k)$ in the time interval $[iT/d, (i+1)T/d]$ (which has duration $T/d$) is $1- \exp(-c'\rho T/d)$. Thus, by the Markov property, the probability of a percolation along this path of length $k$ is at least the product of these terms.

\end{document}